\documentclass[
    reprint,amsmath,amssymb,aps,floatfix
]{revtex4-2}

\usepackage{angles}
\usepackage{dex}

\usepackage[dvipsnames]{xcolor}
\usepackage[polish,english]{babel}
\usepackage[utf8]{inputenc}
\usepackage[T1]{fontenc}
\usepackage{graphicx}
\usepackage[caption=false]{subfig} 
\usepackage{dcolumn}
\usepackage{bm}
\usepackage{hyperref}
\usepackage{amsmath}
\usepackage{amsthm}
\usepackage{amssymb}
\usepackage{multirow}
\usepackage{booktabs}
\usepackage{mathrsfs}
\usepackage{mathtools}
\usepackage{rotating}
\usepackage{setspace}
\usepackage{tikz}
\usepackage{standalone}

\usepackage{microtype} 

\begin{document}
\singlespacing

\theoremstyle{remark}
\newtheorem*{problem}{Problem}

\theoremstyle{plain}
\newtheorem{theorem}{Theorem}
\newtheorem{lemma}[theorem]{Lemma}
\newtheorem{corollary}[theorem]{Corollary}

\title{Visual relativistic mechanics}

\author{Karol Urba\'nski}
\email{karol.j.urbanski@doctoral.uj.edu.pl}
\affiliation{Szkoła Doktorska Nauk Ścisłych i Przyrodniczych\\ Institute of Physics, Jagiellonian University in Kraków.}

\begin{abstract}
    This article shows how to express relativistic concepts in a visual manner using the full power of hyperbolic trigonometric functions. Minkowski diagrams in energy-momentum space are used in conjunction with hyperbolic triangles. Elegant new derivations of the relativistic rocket equation and the relativistic Doppler effect are presented that use this visual approach.
\end{abstract}

\keywords{relativity,accelerated-motion,spacecraft,rocket-equation,kinematics,collisions,rapidity}
\maketitle


\section{Introduction}

Special relativity can be approached in many ways at differing levels of complexity, including using special relativity as a springboard for learning the basics of tensor calculus using four-vectors.  Visualisations using Minkowski diagrams are common. \cite{rowe2009look} These diagrams are an excellent tool for showing worldlines of particles; their light cones encode the causal spacetime structure. One can also easily illustrate Lorentz boosts by `slanting' the $x$ and $t$ axes of the diagram together.

A relative of the Minkowski diagram that provides a visualisation useful in the case of collisions is the Minkowski energy-momentum space diagram. Saletan drew attention to the fact these diagrams are powerful for showcasing the conservation of four-momentum in inelastic scattering and 1+2 dimensional Compton scattering. \cite{saletan1997minkowski} Bokor analyzed collisions in 1+2 dimensional Minkowski space, and calculated the Compton scattering effect by using ordinary geometry and trigonometry on the Minkowski diagram. \cite{bokor2011analysing} In another article, he used ordinary geometry and trigonometry to derive the equation of the relativistic rocket. \cite{bokor2018relativistic} In \cite{paredes2022relativistic}, Paredes et al.\ showed how velocity addition works on such a diagram. In \cite{ogura2018diagrammatic}, Ogura used a diagrammatic tool to analyze elastic collisions. A wonderful, geometric introduction to special relativity that uses these diagrams for calculations has been created by Bais in the book \cite{bais2007very}.

However, the startling thing when looking at this variety of approaches is how the visualisations are separated from the fundamentals of the theory. The aforementioned sources that use visualisation tend to use the Euclidean geometry of the Minkowski graph on paper, instead of the intrinsic geometry of Minkowski spacetime. Sources that emphasize the hyperbolic nature of Minkowski spacetime in their analysis do exist \cite{taylor1992spacetime}; however, they often stick to describing this in algebraic terms, with visualisation in their case feeling like an afterthought. 

A notable exception to this is the recent book by Dray \cite{dray2012geometry}, which treats the basics of special relativity using geometric methods and hyperbolic trigonometry in a manner very similar to this article, and which I highly recommend. However, that work stops short of attempting to use it in more complicated derivations. In this article I will showcase the merits of doing so. It turns out that using rapidity and visual/geometric techniques for performing hyperbolic trigonometry, we can obtain elegant descriptions for complicated physical phenomena.

\subsection*{Notation}
Hyperbolic angles will be represented by $\zeta$ or $\omega$ in this article. Every polar angle used in this article will be marked with the $\theta$ symbol.

The constants $h$ and $c$ are assumed to have a numeric value of $1$, except where indicated explicitly. The metric signature is negative for the temporal dimension and positive for the spatial dimensions.

\subsection{Polar and hyperbolic radians}
In ordinary, Euclidean $2$-dimensional space, distances between points $a = (a_x, a_y) $ and $b = (b_x, b_y)$ are measured using the Pythagorean theorem:
\begin{equation}
    \Delta s^2 = (b_x - a_x)^2 + (b_y - a_y)^2.
\end{equation}
While individual coordinates enter this equation, it is important to note that the equation is invariant with respect to translations and rotations of the original coordinate system. To measure distances along arcs, a quantity known as the line element is useful:
\begin{equation}\label{eq:line-el-euclid}
    ds^2 = dx^2 + dy^2,
\end{equation}
where an element of length along the arc $ds$ is dependent on elements of length $dx$ and $dy$ along the coordinates. This definition can be extended to a higher number of dimensions, and provides an algebraic method of computing distances.

This means that for any point in Euclidean $2$-space, we can find all points at a specified shortest distance $r$ from that point: the result is a circle. In $n$ dimensions, the resulting shape is an $n-1$-dimensional sphere.

\begin{figure}[b]
    \setkeys{Gin}{width=0.48\linewidth}
    \fbox{
        \subfloat[]{\includegraphics{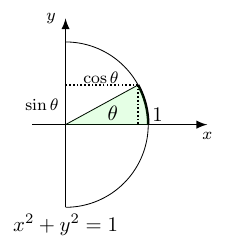}\label{fig:angles}}\hfill%
        \subfloat[]{\includegraphics{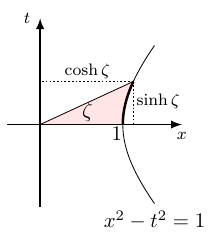}\label{fig:angles2}}
    }
    \caption{The polar angle of $\theta$ radians can be identified with double the area enclosed between the x axis, a ray and a circle, divided by the radius squared to obtain a dimensionless quantity. An analogous definition applies to a hyperbolic angle of $\zeta$ hyperbolic radians, but with a hyperbola replacing the circle. Equivalent definitions using arclengths (marked with a thick line) divided by the radius are also available.}
    \label{fig:hyperbolic-angle}
\end{figure}

Using such a circle, we can formalize a measure of how much one ray originating from the center point needs to be rotated to be coincident with another ray originating from the center point: in figure \ref{fig:angles}, a unit circle $x^2 + y^2 = 1$ is shown with a central angle $\theta$ between the $x$ axis and an arbitrary ray. The value of this angle is obtained by taking the length of the arc (integrated using the line element \eqref{eq:line-el-euclid}), and dividing it by the radius (in this case $1$). However, it can also be integrated in polar coordinates, where an element of length $r d\theta$ is used. By comparing these two quantities, we obtain a definition of the angle in radians.

While this may seem like a convoluted method to define the familiar notion of an angle, it gives a concrete link between geometry and algebra and is easily extensible to hyperbolic geometries. An alternative way to obtain a numerically identical, dimensionless quantity would be to take double the area of the sector of the circle, and divide it by the radius squared.

The trigonometric functions $\cos{\theta}$ and $\sin{\theta}$ are the $x$ and $y$ coordinates on such a unit circle, respectively.

Despite what the ancients thought, Euclidean space is not the only space in which we can do geometry. In Minkowski spacetime with 1 temporal and 1 spatial dimension (known as 1+1 dimensional spacetime), distances are measured using a different line element, called the Lorentzian distance:
\begin{equation}\label{eq:line-el-minkowski}
    ds^2 = dx^2 - dt^2.
\end{equation}
This distance is an invariant with respect to translations, but also with respect to a very important transformation similar to an ordinary rotation: we will take a closer look at that in section \ref{sec:rel-hyp}. In relativistic physics, this distance turns out to be the proper time along an arc connecting two points, called `events': that is, the time measured on a clock carried by an observer moving along a path in spacetime.

Back to our discussion of angles: in direct analogy with the polar angles, we can find all points at a specified minimal distance $s$ from a central point: we will obtain hyperbolas, separated by two asymptotic, diagonal lines. One such unit hyperbola, $x^2 - y^2 = 1$, is shown in figure \ref{fig:angles2}. As in the case of the polar angle, the hyperbolic angle can be defined by taking the length of the arc between two rays and dividing it by the radius; or by doubling the area enclosed by the rays and a hyperbola, and dividing by the radius squared.

We can also define trigonometric functions in this spacetime, called the hyperbolic trigonometric functions, also shown in figure \ref{fig:angles2}.

Minkowski space can have more than 1 spatial dimension. In special relativity, a 1+3 spacetime (with one temporal and three spatial dimensions) is the model we use to describe nature. The temporal dimension is often denoted $t$ (reminder that we have chosen $c = 1$, but one could maintain conventional units by simply relabeling the $t$ axis as $ct$). The others are the spatial dimensions (denoted by $x$, $y$, and so on). When we take a hypersurface of constant time, the space induced on this hypersurface operates like Euclidean space of however many spatial dimensions we originally had.

\subsection*{Useful hyperbolic trigonometry identities}
Since we will be doing calculations using hyperbolic trigonometric functions, here is a refresher on their identities. The functions $\sinh$ and $\cosh$ are the odd and even parts of the exponential function respectively:
\begin{align}
    \sinh{x} =& \frac{e^x - e^{-x}}{2},\label{eq:sinh-e}\\
    \cosh{x} =& \frac{e^x + e^{-x}}{2},\label{eq:cosh-e}
\end{align}
which also means
\begin{align}
    \sinh(-x) =& -\sinh{x},\\
    \cosh(-x) =& \cosh{x}.
\end{align}
Their derivatives are
\begin{align}
    \frac{d\sinh{x}}{dx} =& \cosh{x},\\
    \frac{d\cosh{x}}{dx} =& \sinh{x}.
\end{align}
They satisfy an equation similar to the Pythagorean identity:
\begin{equation}\label{eq:pythagorean-hyperbolic}
    \cosh^2 x - \sinh^2 x = 1,
\end{equation}
and their ratio is
\begin{equation}
    \tanh x = \frac{\sinh x}{\cosh x}.
\end{equation}
Finally, we have three angle addition formulae:
\begin{align}
    \sinh(x + y) =& \sinh{x} \cosh{y} + \cosh{x} \sinh{y},\label{eq:sinh-addition}\\
    \cosh(x + y) =& \sinh{x} \sinh{y} + \cosh{x} \cosh{y},\label{eq:cosh-addition}\\
    \tanh(x + y) =& \frac{\tanh{x} + \tanh{y}}{1 + \tanh{x} \tanh{y}}.\label{eq:tanh-addition}
\end{align}

These identities are very similar to ordinary trigonometric identities. The addition identities also do not have negative signs in front of products, which makes them (arguably) easier to remember.

\section{Relativity and hyperbolic geometry}\label{sec:rel-hyp}
Relativistic mechanics is usually described using the theory of Lorentz transformations, which in $1+3$ dimensions describe six isometries of Minkowski spacetime: three ordinary spatial rotations (in the $XY$, $XZ$, and $YZ$ planes), and three so called `boosts' (in the $X$, $Y$ and $Z$ directions). A boost in the $x$ direction with velocity $v_x$ is described using the following matrix:
\begin{equation}\label{eq:matrix-beta}
    B_x(\beta) = \begin{bmatrix}
    \gamma & -\beta \gamma & 0 & 0\\
    -\beta \gamma & \gamma & 0 & 0\\
    0 & 0 & 1 & 0\\
    0 & 0 & 0 & 1
    \end{bmatrix},
\end{equation}
where $\beta = v_x/c$ is the velocity of the boost and $\gamma = \frac{1}{\sqrt{1-\beta^2}}$ is the Lorentz factor. Matrices of this kind are part of the Lorentz group, which is the set of all the transformations that preserve the speed of light in flat spacetime, a crucial postulate of special relativity. Applying matrix \eqref{eq:matrix-beta} to a Lorentz covariant four-vector in the rest frame will give us the components of the four-vector in the boosted frame. Motion is characterised using the four-velocity 
\begin{equation}\label{eq:four-vel}
    u^\mu = \frac{dx^\mu}{d\tau},
\end{equation}
where $x^\mu$ are components of a worldline parametrized by $\tau$, and the $\tau$ parameter is the proper time measured by a clock carried by our observer. Four-velocity normalises to $u^\mu u_\mu = -1$. Using this, we can show the famous velocity addition formula for collinear motion with velocities $\beta_1$ and $\beta_2$:
\begin{equation}\label{eq:vel-add}
    \beta' = \frac{\beta_1 + \beta_2}{1+\beta_1 \beta_2},
\end{equation}
which tells us that as we get closer to the speed of light, it becomes harder to gain speed. In kinematic descriptions of collisions, we usually use four-momentum vectors:
\begin{equation}
    p^\mu = m u^\mu,
\end{equation}
where $m$ is the rest mass of the object (invariant mass in the object's frame). Using \eqref{eq:four-vel}, we see it normalises to $p^\mu p_\mu = -m^2$. We can describe the total momentum of a collision in an inertial frame by adding four-momenta together:
\begin{equation}
    p^\mu_{\textrm{tot}} = p^\mu_1 + p^\mu_2 + ...
\end{equation}
and four-momentum is conserved in such a collision, meaning:
\begin{equation}
    p^\mu_{\textrm{before}} = p^\mu_{\textrm{after}},
\end{equation}
the application of which has tortured physics undergrads since time immemorial. 

By using the principle of least action and confirming with empirical observation, we find that the time component of four-momentum is the relativistic energy of the particle in the frame, and the relativistic three-momentum is described by the remaining spatial components:
\begin{equation}\label{eq:four-mom}
    p^\mu = (E, \vec{p}) = (\gamma m, \gamma m \vec{v}).
\end{equation}
When we take the norm of the four-momentum, we find the components obey the equation
\begin{equation}\label{eq:e-m-relation}
    (m c^2)^2 = E^2 - (|\vec{p}|c)^2,
\end{equation}
which for massive particles in the rest frame leads to the famous expression of mass-energy equivalence
\begin{equation}\label{eq:m-e-equivalence}
    E = m c^2,
\end{equation}
and for massless particles in any inertial frame to the relation
\begin{equation}\label{eq:p-e-equivalence}
    E = |\vec{p}|c.
\end{equation}

\subsection{Rapidity as the measure of relativistic speed}
Despite the elegance of this approach, some things can remain difficult to a fledgling student. For instance, the formula \eqref{eq:vel-add} of velocity addition is more complicated than one would naively expect in Galilean physics. In addition, the form of the $\gamma$ parameter can be mysterious if care isn't taken to explain where this particular algebraic form comes from. One method we can use to explain its origin is the fact that boosts are well described using the hyperbolic geometry that underlies Minkowski space. A complete derivation of Minkowski spacetime starting from simple axioms a-la-Hilbert can be found in Ref. \cite{schutz1997independent}. I mention this as a curiosity, but it is instructive to understand that one can take axioms based on possible ordering of three events and on some events being unreachable from some other events -- and obtain a geometry described using hyperbolic trigonometric functions that is isomorphic to Minkowski spacetime.

With this in mind, let's look at the geometry with a keener eye. Using the isomorphism established in Ref. \cite{schutz1997independent} we can define the following substitutions: 
\begin{align}
    \gamma =& \cosh{\zeta}, \label{eq:gamma-beta-identities}\\
    \gamma \beta =& \sinh{\zeta},\label{eq:gamma-beta-identities-2}\\
    \beta  =& \tanh{\zeta}.\label{eq:gamma-beta-identities-3}.
\end{align}
These lead to a transformation matrix that fulfills the requirements of being in the Lorentz group: 
\begin{equation}\label{eq:matrix-sinh}
    B_x (\zeta) = \begin{bmatrix}
        \cosh{\zeta} & -\sinh{\zeta} & 0 & 0\\
        -\sinh{\zeta} & \cosh{\zeta} & 0 & 0\\
        0 & 0 & 1 & 0\\
        0 & 0 & 0 & 1
    \end{bmatrix},
\end{equation}
which bears a striking resemblance to a spatial rotation matrix, but with hyperbolic trigonometric functions. For this reason a boost is sometimes called a `hyperbolic rotation' in spacetime. Since this matrix is in the Lorentz group, transformations that use it preserve the speed of light and thus fulfill the postulates of special relativity. The $\zeta$ parameter is customarily called \textbf{rapidity} and in geometrized units is readily interpreted as a hyperbolic angle in dimensionless hyperbolic radians from figure \ref{fig:hyperbolic-angle}. (Note that the Greek letter chosen to represent rapidity even looks like a hyperbola!)

If we derive the velocity addition formula by multiplying two matrices in the form of Eq. \eqref{eq:matrix-sinh}, we will obtain:
\begin{equation}\label{eq:rapidity-addition}
    \zeta' = \zeta_1 + \zeta_2,
\end{equation}
showing that the addition of rapidities in one dimension is linear! The equations \eqref{eq:gamma-beta-identities}, \eqref{eq:gamma-beta-identities-2} and \eqref{eq:gamma-beta-identities-3} give us an alternative, physical definition for the hyperbolic angle between two timelike vectors: since $\gamma$ is the result of contracting four-velocities of two passing or colliding particles, we can measure it using physical means, and use an inverse hyperbolic trigonometric function to get the hyperbolic angle. We can also see why adding regular velocities looked as it does in equation \eqref{eq:vel-add}. In terms of hyperbolic trigonometry (from equation \eqref{eq:tanh-addition}):
\begin{equation}
    \beta' = \tanh(\zeta_1 + \zeta_2) = \frac{\tanh{\zeta_1} + \tanh{\zeta_2}}{1 + \tanh{\zeta_1} \tanh{\zeta_2}} = \frac{\beta_1 + \beta_2}{1 + \beta_1 \beta_2}.
\end{equation}
To drive home that rapidity is a useful concept for relativistic velocity, if we expand $\tanh{\zeta}$ into a series we get
\begin{equation}
    \beta = \tanh{\zeta} = \zeta - \frac{\zeta^3}{3} + \frac{2 \zeta^5}{15} - O(\zeta^7),
\end{equation}
so for low values, rapidity and velocity coincide as in $\beta \approx \zeta$, both dimensionless. For high values -- unlike the speed limit of light equal to $1$ that shows up in \eqref{eq:vel-add} -- rapidity can take any real value, with light being the asymptotic limit at infinity. In this way, the rapidity addition law embodies the physical principle that no matter how many Lorentz boosts we stack together, we can never reach the speed of light.

Using the substitutions \eqref{eq:gamma-beta-identities} and \eqref{eq:gamma-beta-identities-2}, we can rewrite \eqref{eq:four-mom}:
\begin{equation}\label{eq:rel-mom}
    p^\mu = (E, p) = (m \cosh{\zeta}, m \sinh{\zeta}),
\end{equation}
and in doing so we find that the celebrated energy-momentum relation \eqref{eq:e-m-relation} is a physical manifestation of the hyperbolic Pythagorean identity \eqref{eq:pythagorean-hyperbolic}:
\begin{equation}
    E^2 - p^2 = m^2 (\cosh^2{\zeta} - \sinh^2{\zeta}) = m^2.
\end{equation}

\begin{figure}[b]
    \fbox{\includegraphics[width=0.97\linewidth]{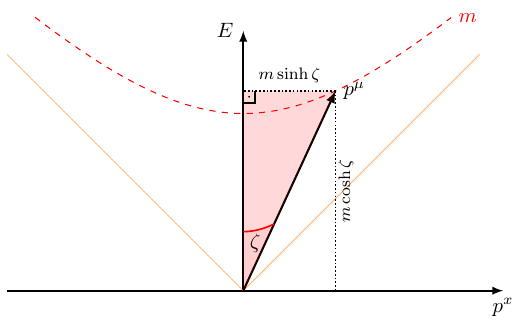}}
\caption{$p^\mu$ is a four-momentum vector of a particle of mass $m$. This represents a particle in a specific spacetime frame of reference. The vector starts at the origin, and ends on a hyperbola of hyperbolic radius $m$; this line is a line of equal rest mass, and the particle's four-momentum must lie on it regardless of reference frame. That is, if we were trying to represent this particle in a frame moving with a different velocity in the $x$ direction, the vector would lie on a different position on the dotted line. Marked is the angle $\zeta$ corresponding to the rapidity of the particle in the frame. Also marked with dotted lines are the values of four-momentum components $E = m \cosh \zeta$ and $p = m \sinh \zeta$, and the light cone. The shaded triangle is an example of a right hyperbolic triangle subtended by the four-momentum vector, the $E$ component and the $p$ component.}
    \label{fig:minkowski}
\end{figure}

\begin{figure}[b]
\fbox{\includegraphics[width=0.97\linewidth]{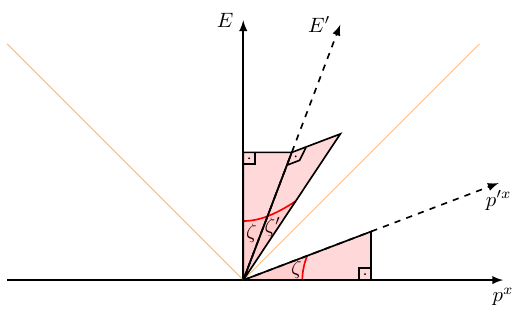}}
\caption{Geometric similarity between triangles on a Minkowski energy-momentum space diagram. The dotted $E'$ and $p'_x$ axes span a frame boosted with rapidity $\zeta$ in relation to the lab frame $E$ and $p_x$. Note that in the boosted frame, even though visually the axes look closer together, they are orthogonal to each other in Minkowski space. The three shaded triangles are all right hyperbolic triangles, with the right angle marked: note the slanted appearance of the marking on a triangle set in the boosted frame with angle $\zeta'$. The triangles have been chosen to be geometrically similar to each other (so $\zeta'=\zeta$), yet possess different physical properties: one is composed of two spacelike and one timelike line, while the other two are composed of two timelike and one spacelike line. The angle $\zeta$ can be found between the $p_x$ and $p'_x$ axes just like it can be found between the $E$ and $E'$ axes, which is a very useful property.}
    \label{fig:boost}
\end{figure}

\subsection{Minkowski diagrams in energy-momentum space}
By now it is clear many relativistic concepts are intimately connected with hyperbolic trigonometry. To visualise these connections, we will use Minkowski diagrams in energy-momentum space. A diagram of this type is a Minkowski diagram in which the axes are $E$ and $p$ instead of $t$ and $x$. We can plot a four-momentum vector on such a diagram, as in figure \ref{fig:minkowski}.

On that diagram, we can spot a right hyperbolic triangle, composed of two timelike lines (the axis $E$ and the four-vector itself) and one spacelike line parallel to the axis $p^x$, meeting at a right angle between a spacelike and timelike line. Between the two timelike lines, we have an angle of $\zeta$ hyperbolic radians. This angle is defined just as in figure \ref{fig:angles2}. This angle parameter has a physical interpretation as a measure of relative velocity between the particle and the resting frame, that we have called rapidity. 

On this triangle, we can use hyperbolic trigonometric functions analogously to ordinary trigonometry: $\sinh \zeta = \textrm{opposite}/\textrm{hypotenuse}$, $\cosh \zeta = \textrm{adjacent}/\textrm{hypotenuse}$, $\tanh \zeta = \textrm{opposite}/\textrm{adjacent}$, and so forth. However, having made the connection between hyperbolic rotations and relative motion in relativistic physics, we can obtain physical insight from our geometric drawings. To wit, the geometric meaning of the expressions \eqref{eq:rel-mom} connecting $E$, $m$ and $p$ is such that the mass $m$ is the ratio of similarity for triangles of particles of differing mass. In an individual triangle, ratios of spacetime lengths behave according to hyperbolic trigonometry. The dashed hyperbola represents all the possible momentum four-vectors for a particle of mass $m$. 

One striking conclusion of this is that the energy $E$ of the particle must be minimal in its own frame -- to observers in relative motion, the energy of a particle is greater. This intersection point of the rest mass $m$ hyperbola and the $E$ axis is the arguably most famous equation in physics -- the mass-energy equivalence \eqref{eq:m-e-equivalence} -- in visual form!

A right hyperbolic triangle can also be composed of two spacelike lines and one timelike line, as long as one of the spacelike and one of the timelike lines are orthogonal to each other (in the sense of their dot product equalling 0).  Examples of various similar right hyperbolic triangles, which allow us to see otherwise hidden hyperbolic trigonometry relations are shown in figure \ref{fig:boost}. This figure also shows a boosted frame $E'$ and $p'$. The boosted frame's behavior is the same as for a traditional Minkowski diagram, where the axes for space and time are brought together. Remember: this changing of axes represents a hyperbolic rotation in Minkowski space, in contrast to a normal rotation in Euclidean space.

Try to understand what's happening in this diagram: Why do right angles appear slanted when we stack two boosts with rapidity $\zeta$ together \footnote{Minkowski diagrams bring axes together, but the axes remain orthogonal. This means a square appears like a parallelogram in our pictorial representation.}? What would be the combined value of rapidity for two such boosts \footnote{In accordance with \eqref{eq:rapidity-addition}, the combined rapidity is $\zeta + \zeta' = 2 \zeta$.}? 

\begin{figure}[b]
    \fbox{\includegraphics[width=0.97\linewidth]{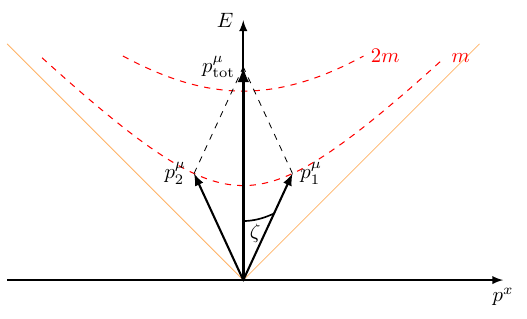}}
    \caption{Four-momentum addition $p_{\textrm{tot}}^\mu = p_1^\mu + p_2^\mu$ using the parallelogram rule. Were this to represent an inelastic collision of two particles of rest mass $m$, we can see mass is not necessarily conserved ($m_{\textrm{tot}} = 2 m \cosh \zeta \geq 2 m$).}
    \label{fig:addition}
\end{figure}

Luckily, despite the different behavior of the axes in the boosted Minkowski diagram, some things continue to work just as we expect in Euclidean space: the four-momentum sum can be constructed using the parallelogram rule. This is shown in figure \ref{fig:addition} for two particles of equal, but opposite momentum in the lab frame. Subtraction of four-momenta also works the same way vector subtraction in Euclidean space does. Parallel lines remain parallel in other frames.

\section{Examples of use: basic operations}
\subsection{Inelastic collision}
An inelastic collision starts with two particles $p_1^\mu$ and $p_2^\mu$, and ends with $p^\mu$. We can add the initial four-momenta using the parallelogram rule and obtain a total four-momentum after collision, as in figure \ref{fig:addition}. 

Taking a closer looks at that diagram -- where momenta were chosen to be equal in magnitude but opposite in direction in the lab frame, and the mass of each particle is equal to $m$ -- we see the final four-momentum does not lie on the hyperbola corresponding to a rest mass of $2m$. This shows an important fact: mass is not necessarily conserved in relativistic mechanics. This is a visual representation of the four-momentum conservation law!

Other examples of such diagrams can be found in articles \cite{bokor2011analysing} and \cite{ogura2018diagrammatic}, as well as Dray's fantastic book \cite{dray2012geometry}.
\begin{figure}
    \fbox{\includegraphics[width=0.97\linewidth]{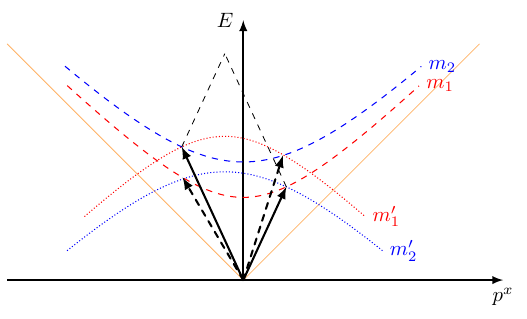}}
    \caption{In an elastic condition in 1+1 dimensions, we can construct possible outcomes by drawing an inverted line of constant rest mass with the origin at the sum of four-momenta. These are represented by the dotted lines. Note that this construction is analogous to using a compass in Euclidean geometry, and is a way to mark vector lengths to accomplish vector subtraction. The constructed line $m'_1$ is drawn such that we could connect any point on the line with the vertex of the parallelogram representing total momentum $p^\mu_{\textrm{tot}} = p^\mu_1 + p^\mu_2$, and obtain a momentum vector that describes a particle with mass $m_1$. Then, the intersection of this line with the line $m_2$ gives the two possible pairs of values $p^\mu_1$ and $p^\mu_2$, one pair of which is the situation before the collision. The remaining pair of vectors must be the only possible outcome after collision, and has been drawn with dashed four-momenta.}
    \label{fig:elastic}
\end{figure}

\subsection{Elastic collision}
An elastic collision starts with two particles $p_1^\mu$ and $p_2^\mu$, and ends with two particles ${p'}_1^{\mu}$ and ${p'}_2^{\mu}$. To represent the conservation of four-momentum, we can add the initial vectors together, just as we did for an inelastic collision. 

After the collision, the particles bounce off each other. Therefore, post collision, we need to have two resulting four-momenta, and they must add to the same conserved value of total four-momentum. The diagrammatic method to do this proceeds as follows: first sum the energy-momentum vectors $p^\mu_1 and p^\mu_2$. Then, from the location of this sum, which represents the conserved energy and momentum, draw an inverted hyperbola representing possible energy and momentum components of particle 1. The locations where this inverted hyperbola crosses the energy-momentum hyperbola of particle 2 gives the particle energy and momentum following the collision. This construction is shown in figure \ref{fig:elastic}, as first seen in Ref. \cite{bokor2011analysing}.

The situation can be extended to $1+2$ dimensions, as was done in that article. With an extra spatial dimension, we now draw an inverted hyperboloid and find the points of intersection with the hyperboloid of the second particle. The possible choices will now be an ellipse, showing that in more than $1+1$ dimensions the deflection angle can be continuously variable, though still constrained.

\subsection{Massless particles}
\begin{figure}
    \fbox{\includegraphics[width=0.97\linewidth]{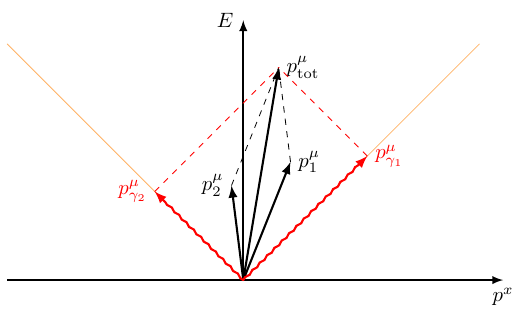}}
    \caption{We can't construct a null vector equalling the center of mass frame's momentum $p^\mu_{tot}$, but we can construct two null vectors $p^\mu_{\gamma 1}$ and $p^\mu_{\gamma 2}$ that combine to it.}
    \label{fig:double-photon}
\end{figure}

We can plot null four-momenta. In that case, in geometric units
\begin{equation}
    p_\gamma^\mu = (E, p) = (E, E) = (\lambda^{-1}, \lambda^{-1}) = (\nu, \nu),
\end{equation}
where $\lambda$ is the wavelength and $\nu$ is the frequency of the photon, meaning the line of constant zero rest mass is the light cone. These vectors have no well-defined rapidity $\zeta$, representing the fact that the speed of light is the unreachable, asymptotic limit, constant in every frame. Nevertheless, while they have no well-defined rapidity, they continue to obey the parallelogram addition rule. We can use this to visualise that two colliding massive particles cannot produce a single photon, but can produce two, as in figure \ref{fig:double-photon}. An interesting application is a derivation of the Compton effect, shown in article \cite{bokor2018relativistic}.

\begin{figure*}
    \setkeys{Gin}{width=0.48\linewidth}
    \fbox{
        \subfloat[]{\includegraphics{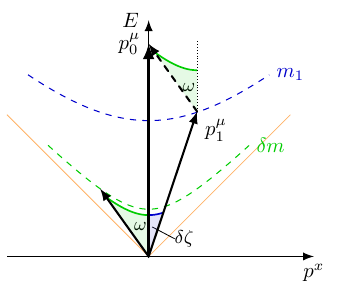}\label{fig:plots:step1}}\hfill%
        \subfloat[]{\includegraphics{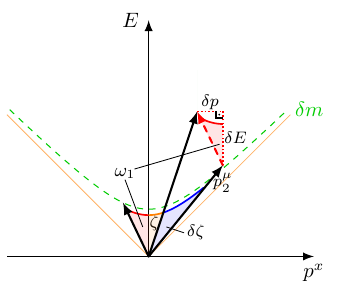}\label{fig:plots:step2}}
    }
    \caption{Momentum space Minkowski diagrams for the relativistic rocket. We showcase the situation assuming large, momentary `impulses'. As the interval between the impulses decreases, we ultimately obtain a smooth curve through energy-momentum space. \textbf{(a)} Plot of first 'step' of acceleration, where we dump $\delta m$ fuel with specific impulse rapidity $\omega$, starting at $p_0^\mu$ at rest in the observer frame, with $p_1^\mu$ being the spacecraft's resulting four-momentum. \textbf{(b)} The situation in a further increment of $\delta m$, where the spacecraft's rapidity changes by $\delta \zeta$. The hyperbolic angle $\omega_1$ is simply the specific impulse $\omega$ added to the current spacecraft rapidity $\zeta$, since rapidities add in hyperbolic space (note the opposite orientation of the angle). Looking at the dotted right triangle, we see $\tanh{\omega_1} = \delta p/\delta E$.}
    \label{fig:relativistic-rocket}
\end{figure*}

\section{Examples of use: advanced derivations}
\subsection{Relativistic rocket equation}
We now turn our attention to something more complicated: the relativistic rocket problem. We start with a spacecraft of mass $m_0$, which expels fuel mass with a given effective exhaust velocity until reaching a dry mass of $m_1$. What will be magnitude of the change of velocity of our spacecraft? In non-relativistic physics, this is usually tackled algebraically using momentum conservation. However, while a similar calculation can be carried out in a relativistic regime, the algebra is more involved, and conservation laws have to be rewritten very carefully; see \cite{forward1995transparent} for the full derivation. We will proceed in a different way, using visual methods.

Initially, we will consider the fuel to be emitted in discrete quantities; in the limit of arbitrarily small quantities, it will correspond to fuel being emitted continuously. Figure \ref{fig:plots:step1} shows the change of momentum when fuel is dumped with effective exhaust rapidity $\omega$ in the rest frame when the rocket starts from rest -- it represents the conservation of four momentum of the rocket/fuel system. Meanwhile, figure \ref{fig:plots:step2} shows the general situation in the original rest frame of the rocket at the instant corresponding to spacecraft rapidity $\zeta$. Because rapidity (unlike velocity) is additive, the rapidity $\omega_1$ with which the reaction mass escapes the exhaust of the spacecraft in this frame is
\begin{equation}
    \omega_1 = \omega + \zeta,
\end{equation}
with angle $\omega$ being negative. The hyperbolic angle $\omega_1$ subtends the marked right hyperbolic triangle, and the adjacent and opposite sides are the absolute magnitudes of the change $\delta E$ and $\delta p$ of the spacecraft (that become $dE$ and $dp$ in the limit of small impulses):
\begin{equation}\label{eq:twosides}
    \tanh{(\omega + \zeta)} = \tanh{\omega_1} = \frac{|\delta p|}{|\delta E|} \rightarrow \frac{|dp|}{|dE|}.
\end{equation}
We rewrite \eqref{eq:twosides}, the left side from definition of the hyperbolic tangent and the right from the definition of $p$ and $E$ in terms of hyperbolic functions of current spacecraft rapidity $\zeta$ and the spacecraft mass $m$:
\begin{equation}\label{eq:twosides-2}
    \frac{\sinh{(\omega + \zeta)}}{\cosh{(\omega + \zeta)}} = \frac{d(m \sinh{\zeta})}{d(m \cosh{\zeta})},
\end{equation}
and use hyperbolic trigonometry and the Leibniz rule to obtain
\begin{equation}
    \frac{\sinh{\omega}\cosh{\zeta} + \cosh{\omega}\sinh{\zeta}}{\cosh{\omega}\cosh{\zeta} + \sinh{\omega}\sinh{\zeta}} = \frac{ m d\zeta \cosh{\zeta} + dm \sinh{\zeta} }{dm \cosh{\zeta} + m d\zeta \sinh{\zeta}}.
\end{equation}
By simple inspection we can glean a possible solution to this equation: \footnote{Actually, we can show that this is also the unique solution. Equation \eqref{eq:twosides-2} can be written as \begin{equation}f\left(\tanh{\omega}\right) = f(m \frac{d\zeta}{dm}),\end{equation} where $f$ is the invertible M{\"o}bius transformation \begin{equation}f(z) = \frac{z \cosh{\zeta} + \sinh{\zeta}}{z \sinh{\zeta}+\cosh{\zeta}}.\end{equation} From this, \eqref{eq:tanh-dzdm} follows directly.}
\begin{equation}
    \begin{cases} \sinh{\omega} = m d\zeta, \\ \cosh{\omega} = dm, \end{cases}
\end{equation}
and the exhaust velocity $\beta_e$ is then by its hyperbolic definition
\begin{equation}\label{eq:tanh-dzdm}
    \beta_e = \tanh{\omega} = \frac{\sinh{\omega}}{\cosh{\omega}} = \frac{m d\zeta}{dm}.
\end{equation}
Using geometric methods, we've obtained an otherwise hidden relationship. This equation is trivial to solve by separation of variables and integration:
\begin{equation}
    |\Delta \zeta| = \beta_e \ln{\frac{m_0}{m_1}},
\end{equation}
which completes the derivation of the relativistic Tsiolkovsky rocket equation in terms of rapidity; $m_0$ is the starting mass and $m_1$ is the mass after acceleration ceases. The standard form of the equation (using ordinary velocity) can be recovered by taking the inverse hyperbolic tangent on both sides. A minus sign is sometimes retained to underscore that the velocity change is directed opposite to the exhaust velocity. 

This derivation is based on the visualisation from article \cite{bokor2018relativistic}, made more elegant thanks to direct usage of hyperbolic trigonometry. The equation in terms of rapidity is also a more natural solution of the problem. If our rocket is multistage, the delta-rapidities add, unlike delta-v:
\begin{equation}
    \Delta \zeta = \sum_{i=1}^{n} \Delta \zeta_i = \beta_1 \ln{\frac{m_0}{m_1}} + \beta_2 \ln{\frac{m_1}{m_2}} + ...
\end{equation}

\subsection{Relativistic Doppler effect: $1+1$ dimensions}
\begin{figure}
    \fbox{\includegraphics[width=0.97\linewidth]{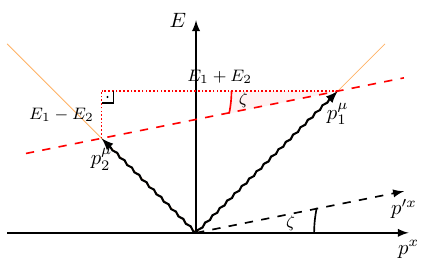}} 
    \caption{Geometric visualisation of the relativistic Doppler effect in one spatial dimension. $p_1^\mu$ and $p_2^\mu$ are four-momentum vectors of massless particles in the receiver's frame. The dashed line connecting the ends of the four-momenta constructs a line parallel to the $p'_x$ axis of the source frame, in which the four-momenta have the same magnitude corresponding to wavelength $\lambda_0$. $\zeta$ thus measures the rapidity of the receiver compared to the source. This angle is found in the dotted right hyperbolic triangle.}
    \label{fig:longitudinal-doppler}
\end{figure}
Geometric methods are especially well adapted to problems where different concepts combine. We will showcase this by deriving the relativistic Doppler effect step by step.

We shall start by restricting ourselves to just one spatial dimension. The source emits omnidirectional electromagnetic radiation of a specific energy corresponding to wavelength $\lambda_0$. In the frame of the source, every emitted photon has a null four-momentum of equal magnitude. We are interested in what happens in the frame of the receiver, boosted with some rapidity $\zeta$ with respect to the source. Therefore, we will draw a situation in that frame, where we expect the four-momenta in opposite directions to differ. This is shown in figure \ref{fig:longitudinal-doppler}: we draw the four-momenta $p_1^\mu$ and $p_2^\mu$ of the photons.

Then, we can ask ourselves what the source frame looked like. This method of looking at a frame from a different point of view is at the heart of relativity. The source frame turns out to be easily constructible: drawing a line through the ends of the four-momenta and taking the $p^x$ axis as parallel to it, we obtain a frame where the momenta have the same magnitude by construction. This construction is shown with the dashed lines in figure \ref{fig:longitudinal-doppler}. The boost thus constructed corresponds to $\zeta$.

\begin{figure*}
    \setkeys{Gin}{width=0.48\linewidth}
    \fbox{
        \subfloat[]{\includegraphics{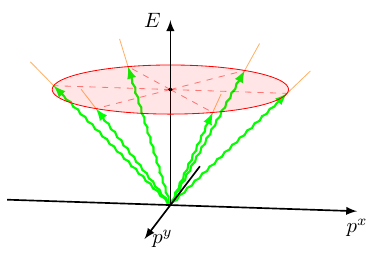}\label{fig:searchlight-source}}\hfill%
        \subfloat[]{\includegraphics{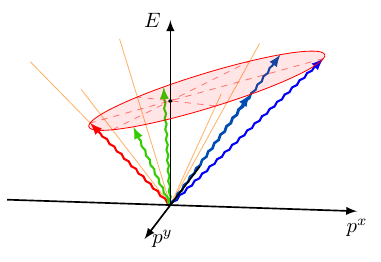}\label{fig:searchlight-receiver}} %
    }
    \caption{Geometric visualisation of the searchlight effect. \textbf{(a)} The light cone in the source frame is cut with a plane to produce a circle of momenta that are equal in magnitude. This corresponds to omnidirectional emission at a certain wavelength. The marked four-momenta are spaced by equal intervals of $\frac{\pi}{3}$, marked on the cutting plane with the dashed lines. \textbf{(b)} In the boosted frame, the plane from the previous picture must remain a plane, but becomes `slanted' in the diagram. The light cone doesn't change shape, in accordance with the constancy of the speed of light in all frames. The cutting plane and the light cone therefore create a conic section: specifically, an ellipse. This `stretching' of the circle into the ellipse moves the ends of the vectors; we can see that they are no longer spaced equally as the marked intervals of $\frac{\pi}{3}$, and have instead moved ahead. Therefore, vectors concentrate towards the direction of travel: the radiation is no longer of the same intensity in every direction. Instead, the faster the boost, the more collimated the beam becomes, like a searchlight. The lengths of the vectors have also changed, and light becomes more blue-shifted towards the direction of travel. In the opposite direction, the rays become red-shifted, though there are fewer and fewer of them in a given angle.}
    \label{fig:general-searchlight}
\end{figure*}

This angle turns out to be possible to also find in the hyperbolic triangle shown with dotted lines. The triangle is a right hyperbolic triangle in the receiver's frame. The marked $\zeta$ is by hyperbolic trigonometry the hyperbolic tangent of the opposite side and the adjacent side. These sides have well defined lengths, since for a null vector in Minkowski 2-space, the temporal and spatial components are exactly equal, and correspond to the energies of the photon. Therefore:
\begin{equation}
    \tanh{\zeta} = \frac{E_1 - E_2}{E_1+E_2},
\end{equation}
which we rewrite on the left from hyperbolic trigonometry and the right from the energy to wavelength relation:
\begin{equation}
    \frac{\sinh{\zeta}}{\cosh{\zeta}} = \frac{\lambda_1^{-1} - \lambda_2^{-1}}{\lambda_1^{-1} + \lambda_2^{-1}}, 
\end{equation}
and finally from the trigonometric identities \eqref{eq:sinh-e} and \eqref{eq:cosh-e}, and basic algebra:
\begin{equation}\label{eq:doppler-intermediate}
    \frac{e^{\zeta} - e^{-\zeta}}{e^{\zeta} + e^{-\zeta}} = \frac{\lambda_2 - \lambda_1}{\lambda_2 + \lambda_1}.
\end{equation}
By inspection it follows that one possible solution \footnote{Once again, we can show this with more rigor and show uniqueness using the invertible M{\"o}bius transformation.} is
\begin{equation}\label{eq:doppler}
    \begin{cases}
        \lambda_1 = e^{-\zeta},\\
        \lambda_2 = e^{\zeta}.
    \end{cases}
\end{equation}
Let's confirm we've got a well-behaved answer. If we take the source frame ($\zeta = 0$) in \eqref{eq:doppler}, the photons have equal wavelength (as indeed they should) $e^{0} = 1$. The original $\lambda_0$ is a geometric unit of the construction and has to be restored:
\begin{equation}\label{eq:doppler2}
    \lambda = \lambda_0 e^{\pm \zeta}.
\end{equation}

Thinking about this from the perspective of the receiver, the photons which hit head-on have higher energy, while the ones that have to catch up have lower energy.

Let's bring the solution to a different form. Since $e^\zeta$ is composed of $\sinh$ and $\cosh$, through the use of \eqref{eq:gamma-beta-identities} and \eqref{eq:gamma-beta-identities-2}:
\begin{equation}
    \lambda = e^{\zeta}\lambda_0 = \lambda_0 (\sinh \zeta + \cosh \zeta) = \lambda_0 \sqrt{\frac{1+\beta}{1-\beta}},
\end{equation}
and we recover the standard form of the longitudinal Doppler shift. 
\subsection{Relativistic Doppler effect: $1+2$ dimensions}
\begin{figure}[b]
    \fbox{\includegraphics[width=0.97\linewidth]{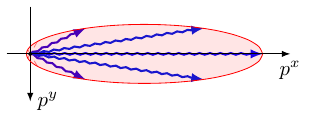}}
    \caption{In a highly relativistic case, the ellipse becomes so elongated, nearly all the light is received by the moving observer from the direction towards which it moves in relation with the stationary source. From the other perspective, a moving source appears to emit most of its radiation towards the direction of its motion by a stationary observer.}
    \label{fig:searchlight-extremal}
\end{figure}

But let's not stop there; the longitudinal effect is the least interesting part of the full Doppler effect. Let's see if we can tackle adding a second spatial dimension. We shall begin with a humble goal of understanding the geometry qualitatively.

Once again, we want to see what happens in the receiver's frame. But with an extra dimension in the way, we will start with a careful examination of how the situation looks in the source frame. This is shown in figure \ref{fig:searchlight-source}: the four-momenta representing omnidirectional emission with equal intensity and equal energy is represented by a circle of four-momenta going in each spatial direction. We can visualise it as a plane `chopping' the light cone to get a circle.

\begin{figure*}
    \setkeys{Gin}{width=0.48\linewidth}
    \fbox{
        \subfloat[]{\includegraphics{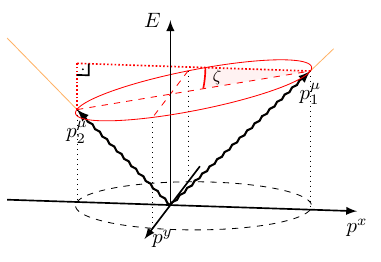}\label{fig:doppler-2d-1}}\hfill%
        \subfloat[]{\includegraphics{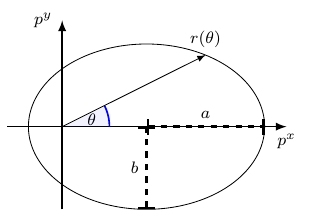}\label{fig:doppler-2d-1-projection}}
    }
    \caption{Geometric visualisation of the relativistic Doppler effect in two spatial dimensions. \textbf{(a)} The light cone in the receiver's frame is cut by the plane of the source spatial axes to produce a conic section; the circle (which represents isotropic emission) becomes an ellipse (anisotropic reception). \textbf{(b)} On the projection of that ellipse to the momentum plane, we see the vector $r(\theta)$ that is proportional to $\lambda^{-1}$ can be described using the angle $\theta$ and the properties of an ellipse centered at one focus.}
    \label{fig:general-doppler}
\end{figure*}

Now, let's imagine this same situation in a frame boosted along the direction of the $x$ axis. We know that this boost can do nothing to change the width of our circle when measured along the $y$ direction, since no boosting is happening along that axis. But, we know from the longitudinal Doppler effect that our circle must change size in the $x$ direction. Therefore, it can no longer be a circle: it must be a different shape. Using our visualisation of a plane cutting the light cone, we can draw this plane very similarly to how we did it for the longitudinal effect: the plane will appear slanted at a certain angle. But, the light cone remains a cone: as we've mentioned, this is a representation of the constancy of the speed of light. Therefore, we obtain a conic section with a slanted plane: an ellipse. See the illustration in figure \ref{fig:searchlight-receiver}.

On both the figures in \ref{fig:general-searchlight}, a sample of four-momentum vectors have been marked. In the first picture, the ends of the marked vectors are equidistant along the arc of the circle. During our boost, the ends of the vectors have to move forwards, to accommodate the change of our shape. But, the origin of the ellipse -- representing the observer -- remains somewhere on the $E$ axis. This means that the four-momenta get stretched; more of them start pointing forwards rather than backwards. The ones pointing forwards get blueshifted, while the ones behind get redshifted. It appears that our light concentrated, like a searchlight!

This concentration of light is called the relativistic beaming effect or the searchlight effect. Objects in rapid motion emit more light towards their direction of motion. When frames are looked at in reverse order (the source is now a receiver), in accordance with the principle of relativity, light received by a moving spacecraft also appears to be concentrated -- when we move rapidly, we receive most of the light from the direction we travel to! It is as if the universe was being squeezed into a viewing window in front of us. \footnote{It appears the warp effect in Star Wars was a lie. How disappointing.} With figure \ref{fig:general-searchlight}, we were able to grasp what physically happens in a situation where multiple effects combine, without doing any algebra. Figure \ref{fig:searchlight-extremal} showcases a highly relativistic instance of the same construction, where the searchlight effect is more dramatic.

But physicists need to be able to measure things and make predictions. So, with our newfound understanding, let's extract an equation from our pictures. Figure \ref{fig:doppler-2d-1} shows the ellipse, but this time, we've marked the same construction we've used in the longitudinal effect, to construct the rapidity of our boost $\zeta$. What interests us are the magnitudes of the momentum vectors: figure \ref{fig:doppler-2d-1-projection} shows the projection of this ellipse on the momentum 2D space. Its focus, by properties of conic sections, must lie at the origin i.e., at $p = (0,0)$. The length $r$ is the momentum of the light signal in direction $\theta$ of the receiver's frame, and is equal to $\lambda^{-1}$.

All that we need to do to derive the necessary formula, then, is to calculate the equation of this ellipse. The semi-minor axis is $b = 1$ in the units of the construction, since the initial width of $2\lambda^{-1}_0$ is unaffected by the boost. The semi-major axis is half the sum of the momenta calculated for longitudinal Doppler shift:
\begin{equation}
    a = \frac{\lambda_1^{-1} + \lambda_2^{-1}}{2} = \frac{e^{-\zeta} + e^{\zeta}}{2} = \cosh{\zeta}.
\end{equation}
The semi-latus rectum (the geometric distance from the focus to the point on the ellipse perpendicular to the semi-major axis) is
\begin{equation} 
    \ell = \frac{b^2}{a} = \frac{1}{\cosh{\zeta}},
\end{equation}
and the eccentricity is
\begin{equation}
\begin{split}
    e =\sqrt{1 - \frac{b^2}{a^2}} = \sqrt{1 - \frac{1}{\cosh^2{\zeta}}} =& \sqrt{\frac{\cosh^2{\zeta} - 1}{\cosh^2{\zeta}}} = \\ = \sqrt{\frac{\sinh^2{\zeta}}{\cosh^2{\zeta}}} =& \tanh{\zeta}.
\end{split}
\end{equation}
Inserting these ellipse properties into the polar equation for the ellipse centered at the origin:
\begin{equation}
    r(\theta) = \frac{\ell}{1 + e \cos{\theta}} = \frac{1}{\cosh{\zeta} \cdot (1 + \tanh{\zeta} \cdot \cos{\theta})},
\end{equation}
we obtain an expression for the relativistic Doppler effect in arbitrary direction measured in the receiver frame. We can confirm this by reestablishing the unit $\lambda_0$ and using hyperbolic definitions of $\gamma$ \eqref{eq:gamma-beta-identities} and $\beta$ \eqref{eq:gamma-beta-identities-3}:
\begin{equation}
    \lambda^{-1} = \frac{\lambda_0^{-1}}{\gamma(1 + \beta \cos{\theta})},
\end{equation}
which is the standard expression found in the literature.

We also see that the parameters of this ellipse are simple in terms of $\zeta$. The semi-latus rectum $\ell$ is, by definition, the geometric distance from the focus to the point perpendicular to the semi-major axis. Thus, once we restore the unit $\lambda_0$, it is the transverse relativistic Doppler effect:
\begin{equation}
    \lambda_T = \lambda_0 \cosh{\zeta}  = \gamma \lambda_0.
\end{equation}
Other parameters are intimately connected with physical properties we use in the ordinary algebraic approaches: the eccentricity is actually the velocity $\beta$, the semi-major axis is $\gamma$, and so forth.

\section{Discussion and pedagogical notes}
My intention when presenting these visual, geometric ideas is not to insist that standard methods are obsolete. On the contrary, the algebra of four-vectors in the Minkowski metric is an excellent introduction to the mathematics that students will use in general relativity. However, the algebra should not displace the geometry entirely. Visualisation methods are a great introduction to thinking about symmetries, a crucial skill.

Maintaining proper balance would be prudent. The chief challenge faced when introducing such methods is that geometric arguments that aren't understandable are frustrating. It is extra cognitive load both for the student and the teacher.

Yet, there are many advantages to this visual style. Concepts such as relativistic mass and velocity addition are readily understood as manifestations of the spacetime geometry. The hyperbolic trigonometric function expressions and identities are very easy to recall when forgotten, as they are similar to familiar trigonometry.

Once a geometric argument is understood, it becomes very intuitive and applicable in other contexts. A feeling of having uncovered a deeper truth often accompanies this understanding. Geometric methods and arguments can showcase a beauty of mathematics different from, and complementary to that of ordinary algebraic methods. This is an excellent motivator for people who enjoy the creative, artful side of mathematics and physics.

In my opinion, when tutoring small classes, visualisation can be developed along with students. In a bigger classroom one could adopt a hybrid approach, where the focus is on traditional methods -- but is then bolstered by presenting the geometry. To this end, the book \cite{dray2012geometry} by Dray expands on the methods shown in the introduction.

The economy of the arguments presented here ensures that they can be shown and explained quickly, or in supplementary material. This also makes them viable for student organised or weekly institutional seminars. 

\section*{Acknowledgements}
I would like to thank Tristan Needham for writing \textit{Visual differential geometry and forms} \cite{needham2021vdgaf}, which inspired my search for deeper geometric meaning and beauty when teaching my special relativity classes, as well as the title and style of this article.

I would also like to thank Sebastian Szybka for pedagogical help, and for suggestions of improvements to this article.

The drawings used code from \cite{tikzsource} as a starting point.

\bibliography{main}
\end{document}